\documentclass[11pt,a4paper]{article}
\usepackage{jinstpub}
\usepackage{graphicx}
\usepackage{newtxtext,newtxmath}
\usepackage[polish,british]{babel}

\title{Multichannel FPGA based MVT system for high precision time (20~ps~RMS) and charge measurement.}

\author{M.~Pa\l ka{$^a$}\thanks{Corresponding author.}, 
P.~Strzempek{$^a$}, 
G.~Korcyl{$^a$}, 
T.~Bednarski{$^a$}, 
Sz.~Nied\'zwiecki{$^a$}, 
P.~Bia\l as{$^a$}, 
E.~Czerwi\'nski{$^a$}, 
K.~Dulski{$^a$}, 
A.~Gajos{$^a$}, 
B.~G\l owacz{$^a$}, 
M.~Gorgol{$^c$}, 
B.~Jasi\'nska{$^c$}, 
D.~Kami\'nska{$^a$}, 
M.Kajetanowicz{$^a$}, 
P.~Kowalski{$^b$}, 
T.~Kozik{$^a$}, 
W.~Krzemie\'n{$^d$}, 
E.~Kubicz{$^a$}, 
M.~Mohhamed{$^a$}, 
L.~Raczy\'nski{$^b$}, 
Z.~Rudy{$^a$}, 
O.~Rundel{$^a$}, 
P.~Salabura{$^a$}, 
N.G.~Sharma{$^a$}, 
M.~Silarski{$^a$}, 
J.~Smyrski{$^a$}, 
A.~Strzelecki{$^a$}, 
A.~Wieczorek{$^a$}, 
W.~Wi\'slicki{$^b$}, 
M.~Zieli\'nski{$^a$}, 
B.~Zgardzi\'nska{$^c$} and 
P.~Moskal{$^a$}\\

{$^a$}Faculty of Physics, Astronomy and Applied Computer Science, Jagiellonian University,\\ 
30-348 Cracow, Poland\\
{$^b$}\'Swierk Computing Centre, National Centre for Nuclear Research,\\
05-400 Otwock-\'Swierk, Poland\\
{$^c$}Department of Nuclear Methods, Institute of Physics Maria Curie-Sk\l odowska University,\\
Pl. M. Curie-Sk\l odowskiej 1, 20-031 Lublin, Poland\\
{$^d$}High Energy Physics Division, National Centre for Nuclear Research,\\
Soltana 7, 05-400 Otwock-\'Swierk, Poland\\

\emailAdd{marek.palka@cern.ch}}

\abstract{
In this article it is presented an FPGA based  
\textbf{M}ulti-\textbf{V}oltage \textbf{T}hreshold (MVT)
system which allows of sampling fast signals ($1-2$ ns rising and falling edge) in both voltage and time domain.
It is possible to achieve a precision of time measurement of $20$ ps RMS and reconstruct charge of signals, using a simple approach, with deviation from real value smaller than 10\%.  
Utilization of the differential inputs of an FPGA chip 
as comparators together with an implementation of a TDC inside an FPGA
allowed us to achieve a compact multi-channel
system characterized by low power consumption and low production costs.
This paper describes realization and functioning of the system comprising
192-channel TDC board and a four mezzanine cards 
which split incoming signals and discriminate them. 
The boards have been used to validate a newly developed Time-of-Flight Positron 
Emission Tomography system based on plastic scintillators. 
The achieved full system time resolution of $\sigma$(TOF) $\approx 68$ ps
is by factor of two better with respect to the current TOF-PET systems. 
}


\begin{document}
%\begin{otherlanguage}{british}
\maketitle

\section{Introduction}

Development of more precise time and charge measurement methods was always pushed forward due to new requirements set by a necessity of building more demanding systems for more sophisticated experiments or devices.
In a standard of precise time and charge measurement approach a~set of preamplifiers, 
comparator chips, \textbf{T}ime to \textbf{D}igital \textbf{C}onverters (TDC), 
ASICs and a separate readout system are used 
to construct the whole measurement system. Typically
readout system is based on the \textbf{F}ield-\textbf{P}rogrammable \textbf{G}ate \textbf{A}rray (FPGA) devices \cite{other_mvt}. 
Such approach leads to complex and relatively large systems 
where a~high density of measurement channels is hard to achieve. 

The first step to merge parts of this system was already made few years ago, when TDC implementation was performed inside an FPGA device by using adders carry chains as delay lines~\cite{tdcpl}. 
Later this method was improved and over time it was possible to achieve 
$20$ ps RMS \cite{tdcwu}. 
The next step, to compactify measurement systems, was encapsulating an ADC into FPGA using its 
\textbf{L}ow \textbf{V}oltage \textbf{D}ifferential \textbf{S}ignalling 
(LVDS) buffers as comparators \cite{adcwu}. This ADC implementation was used for sampling slow signals. Further on, MVT device based almost solely on the FPGA was proposed \cite{myart,mvtold} where in the latter one a coincidence timing resolution of $684$ ps FWHM was achieved. 
It is worth noting that using an FPGA for MVT improves the compactness of the systems.
Additionally it enables to incorporate in the same FPGA real time algorithms 
for specific system requirements. Also it results in less power consumption and finally reduces costs.

Following sections contain description of system components (sec. II), general idea of charge and time measurements (sec. III), characteristics of FPGA LVDS buffers used as comparators (sec. IV) and results achieved with a prototype of \textbf{J}agiellonian \textbf{P}ositron \textbf{E}mission \textbf{T}omograph (J-PET) detector (sec. V). 

In order to have the MVT FPGA based system under full control it was necessary to understand and check feasibility of usage of the FPGA LVDS buffers as a comparators in such constraints defined by the TOF-PET application.

\section{MVT FPGA based electronics}
A core of presented system consists of two electronic boards. The first one is a base board, so called \textbf{T}DC \textbf{R}eadout \textbf{B}oard (TRB), with the TDC implementation in the FPGA. It is described in details in \cite{trb_ref} and its architecture foresees additionally a possibility to connect extension boards for dedicated measurements \cite{trb_app}. The second board, designed by our group, extends the system functionality to serve as an MVT measurement device.

\subsection{TRB3 - TDC implementation}

\begin{figure}[!t]
\centering
\includegraphics[width=0.6\textwidth]{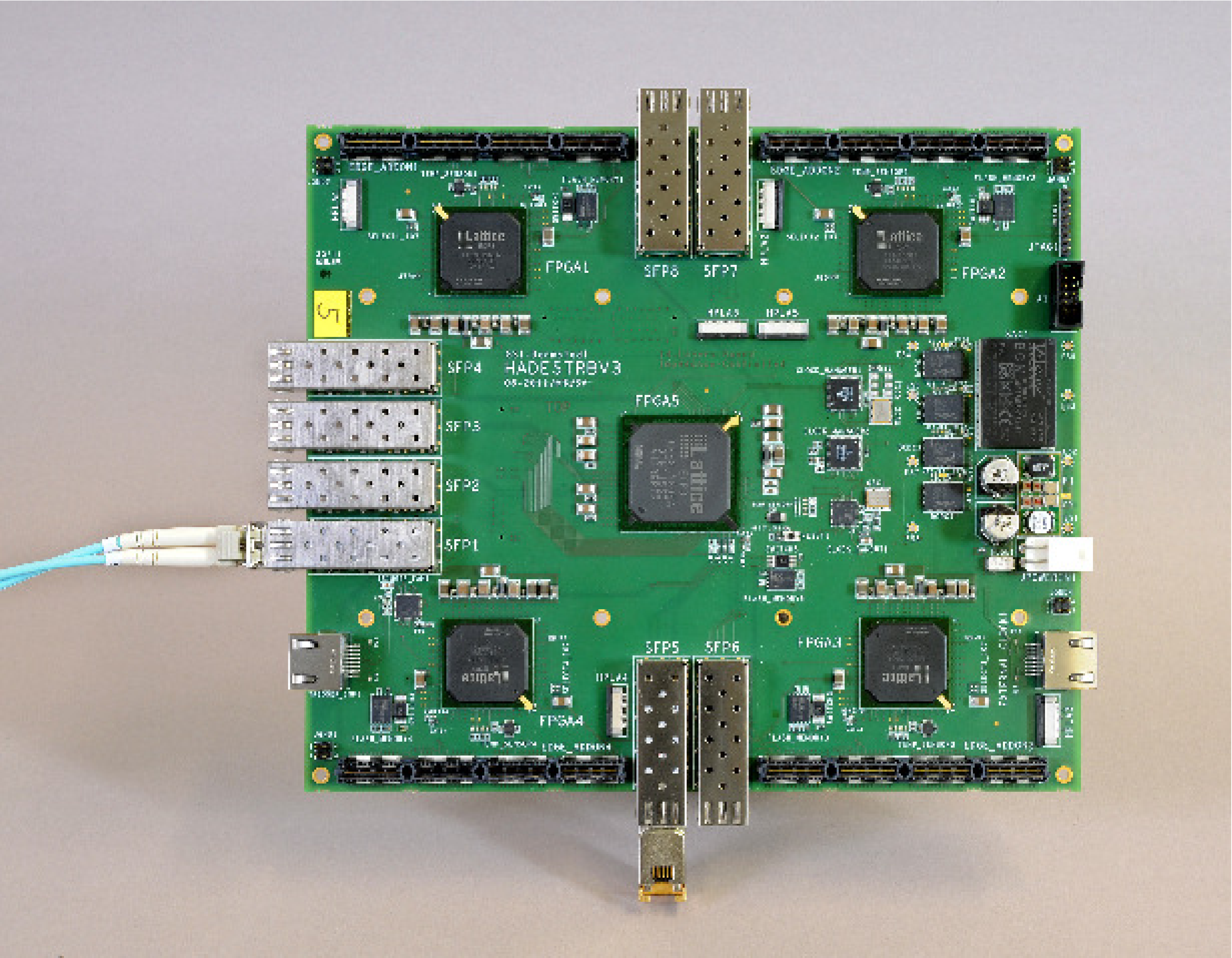}
\caption{The \textbf{T}DC \textbf{R}eadout \textbf{B}oard (TRB) consists of the control FPGA placed in the centre and other four FPGA's  providing 196 TDC channels. On the edges of the board high data rate connectors are located to directly measure signals or to attach mezzanine cards.}
\label{trbv3}
\end{figure}

The TRB board (see Figure \ref{trbv3}) consists of five Lattice ECP3 FPGA units. 
The central FPGA manages data flow on the board for both readout and configuration data.
The other four edge FPGAs provide 196 TDCs channels - rising and falling edge. 
The TRB design together with implemented firmware assures precision of time measurements below $14$ ps RMS \cite{trb_ref}. 
The input signals are expected to be in the LVDS standard. 
The board can be synchronised with other systems by means of measuring one common reference time. In this way it is possible to build 
vast and high channel density systems.

\subsection{MVT front end electronics}
\begin{figure}[!ht]
\centering
\includegraphics[width=0.6\textwidth]{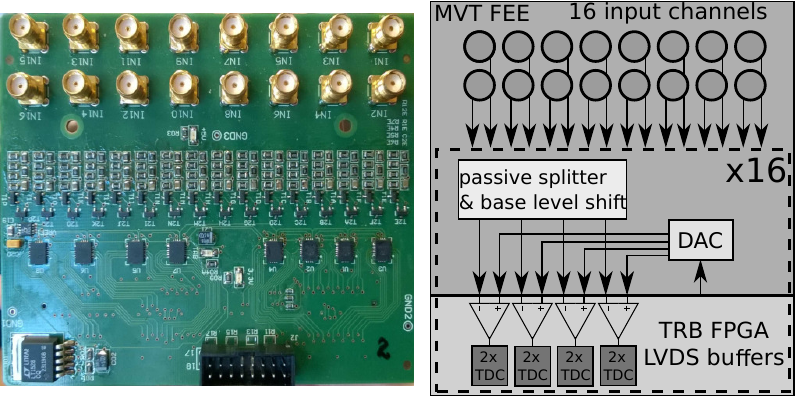}
\caption{Picture of the MVT mezzanine card (left) and block diagram which represents 
its functionality (right). The MVT board has 16 channels where analogue signals are split into four with passive splitters and together with threshold voltage signals generated by DAC are passed to the FPGA LVDS buffers.}
\label{fee_pic}
\end{figure}

The MVT electronics incorporate DAC chips LTC2620 from Linear Technology for threshold settings
and passive splitters, which divide incoming analogue signals into four, 
assuring possibility of applying four independent thresholds. 
The scheme of the MVT mezzanine board is shown in Figure \ref{fee_pic}. 
The DAC outputs and passively split signals are directly connected to the FPGA LVDS buffers which act as comparators. 
It is simple circuit and follows the idea of having most of MVT functionality inside the FPGA device. 

\section{Time and charge measurement}

\begin{figure}[!ht]
\centering
\includegraphics[width=0.8\textwidth]{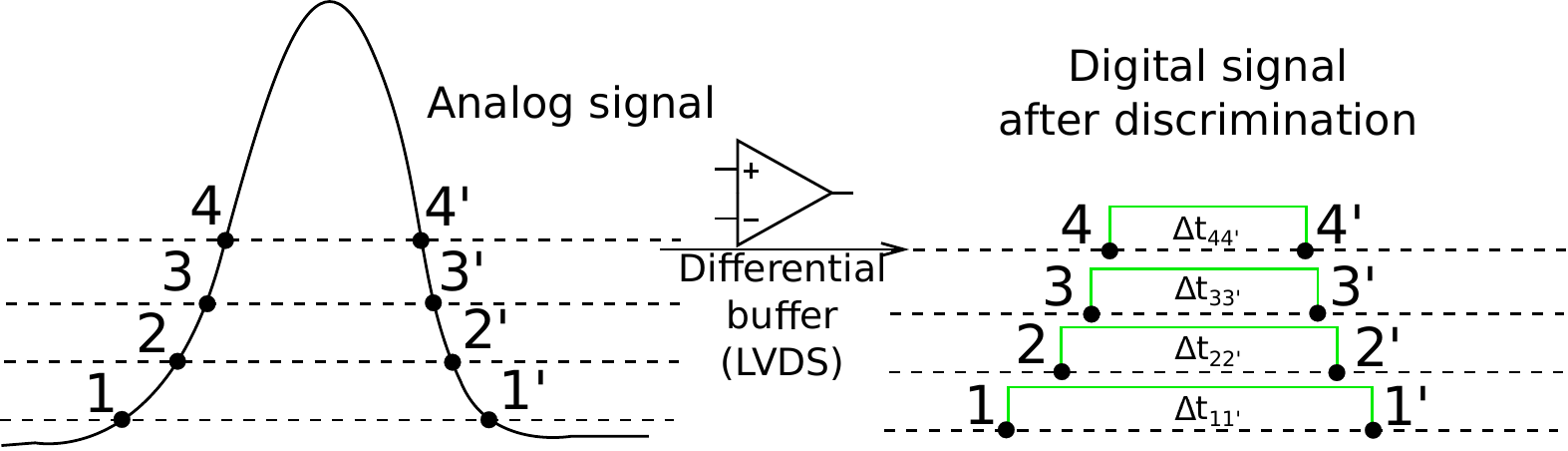}
\caption{The method of signal sampling described in this article. Sampling on different voltage levels allows to determine more precisely the start time of the signal and its charge. It may be done by fitting a curve which describes the shape of the signal using either the method of library of synchronised model signals \cite{JPET_nim2015} or by more advanced methods as e.g. the signal shape reconstruction by means of the compressive sensing theory \cite{res_to_prec,lech_nim_2}.}
\label{tdc_mult_fpga}
\end{figure}

The electronics enables sampling of a measured signal and hence facilitates its reconstruction
with a relatively high accuracy by application of the compressing sensing theory \cite{res_to_prec,lech_nim_2}. 
The thresholds for the measurements have to be adjusted with dedicated DAC's 
to cover the voltage range spanned from the base line level to a maximum signal amplitude (see Figure \ref{tdc_mult_fpga}). 
The comparison of the predefined threshold with the incoming signal is performed by means of FPGA LVDS buffers, 
described in details in the next section. Once the signal crosses the threshold a logical signal inside an FPGA 
is changing and the corresponding time is measured in TDC. 
The time determined from the crossing of the lowest threshold (1) allows to estimate a start time of the signal. 
The times measured at higher thresholds (2,3,4) may be used to improve the precision of the start time determination
e.g. by the reconstruction of the full signal waveform \cite{lech_nim_2} or by a fit of a proper %polynomial 
function to measured points on the rising edge of the signal (either in real time in the FPGA or in the off-line analysis).

\begin{figure}[!ht]
\centering
\includegraphics[width=0.8\textwidth]{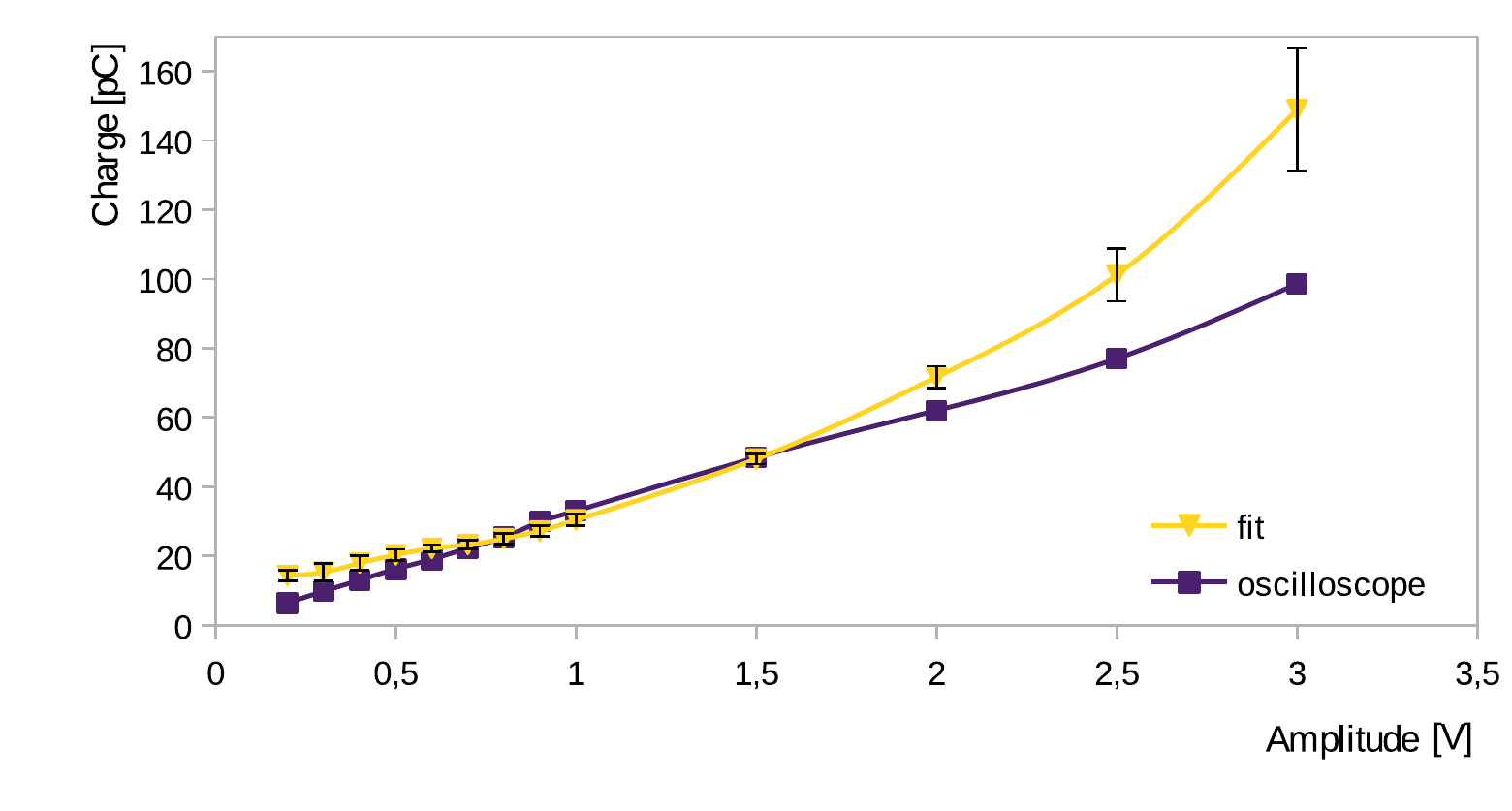}
\caption{The charge of the signal as a function of the amplitude.
Triangles denote the charge (before splitting) reconstructed using the method described in the text,
and squares indicate the charge measured using the Serial Data Analyzer (Lecroy SDA6000A).} 

\label{charge}
\end{figure}

To demonstrate the capability of the charge measurement of fast signals a test signal from generator 
was sampled both by the oscilloscope with time interval of $50$ ps (Lecroy SDA6000A) and by the presented system. 
Signal had $0.9$ ns rise and fall time and was kept for $0.7$ ns high. 
Its amplitude was changed from $0.2$ V to $3$ V. Since signal was split into four and was terminated with $50$~$\Omega$ its signal amplitude on the LVDS buffers was around $6$ times smaller and it was ranging from $0.03$ to $0.5$ V. The corresponding DAC thresholds were set to $37$, $67$, $97$, $127$ mV. The smallest signal (0.03V) is still visible due to the intrinsic FPGA LVDS buffer character. The threshold value set in DAC is seen in the FPGA LVDS buffer with up to ~20mV shift. Influence of this shift can be calibrated for the final system.
In order to establish a~reference value the area under a signal, translated then into charge, was measured with the oscilloscope (see Figure \ref{charge} - squares). 
Next the signal of the same shape was sampled by the MVT-TRB system 
and a Gaussian function was fit to measured points (time and corresponding threshold) and again the charge was calculated (Figure \ref{charge} - triangles). 
Uncertainties of this measurements were calculated as the standard deviation of the Gaussian fit to the measured signal. They varied between $3$ and $7$ $\%$ except that for small ($<0.5$ V) and high ($>3$ V) amplitudes it reached $16$ $\%$. 
It has to be emphasised that for this simplified method there is still a room to reduce the uncertainty of charge measurement by the optimization of the threshold values and better choice of the fit function which would properly describe the shape of the measured signals. 

The development and the choice of the method are beyond the scope of this article,
and the interested reader is referred e.g. to reference \cite{lech_nim_2}.
Here, the most important was to demonstrate that the reconstructed charge follows the real charge measured with the oscilloscope with a decent precision.

\section{Differential buffer characteristics}

\begin{figure}[!t]
\centering
\includegraphics[width=0.8\textwidth]{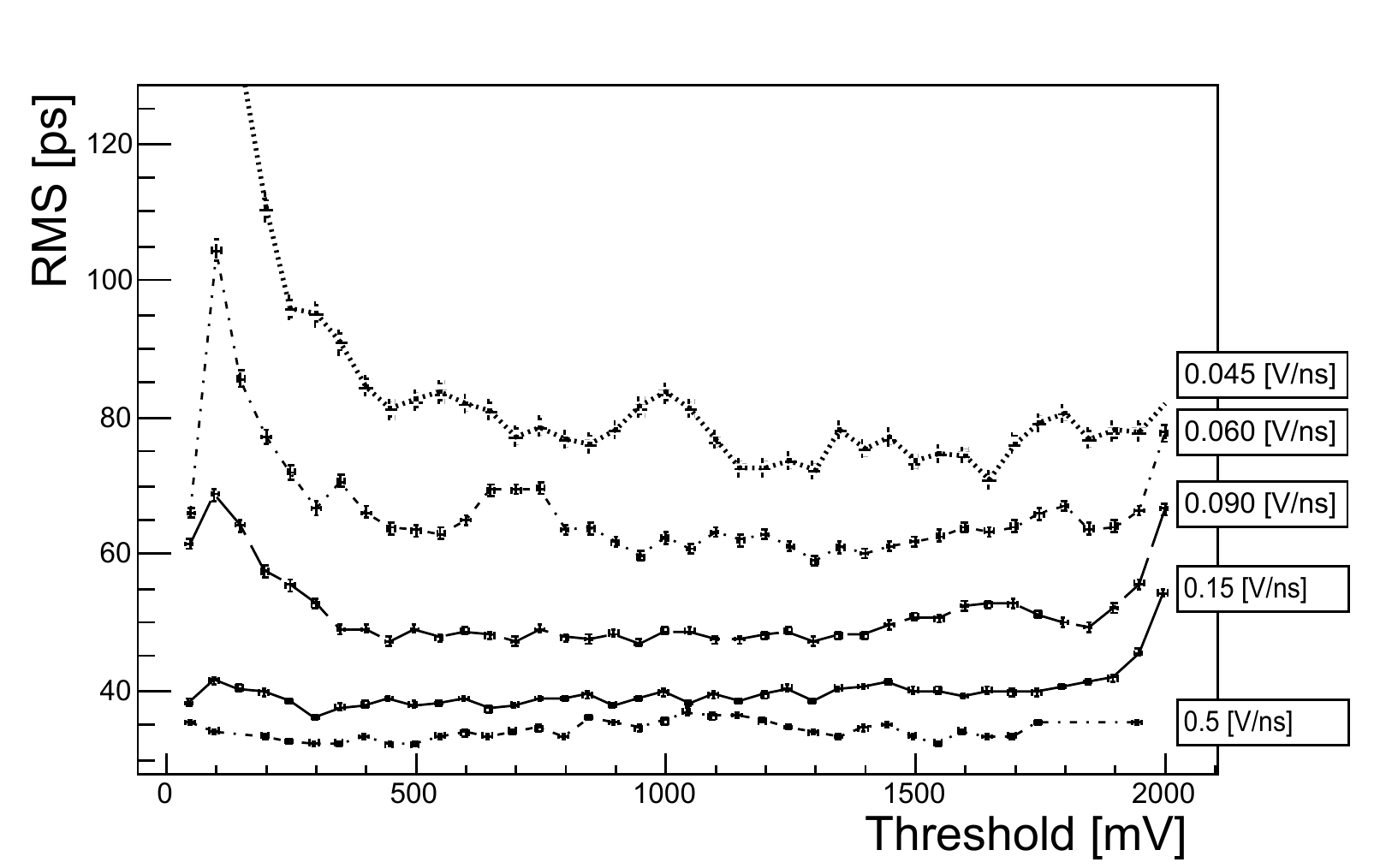}
\caption{The average precision of the time difference measurement as a function of a threshold level and slew rate.}
\label{tdc_v_slope}
\end{figure}

In ECP3 Lattice FPGAs an LVDS buffer (work in a voltage range from $0$ to $2$ V) 
is used normally for a data transfer. Therefore, in order to check if it is possible to use these buffers 
as analogue comparators a set of measurements were carried out for various voltages and slew rates. 
The results, shown in Figure \ref{tdc_v_slope}, represent achieved time measurement precision (RMS) 
of the time difference between two channels as a function of a threshold 
level for pulses with various slew rates.

As it can be seen when slew rate is decreasing the time measurement 
precision worsens. This kind of behaviour is expected since for slower signals 
it is more uncertain when the LVDS buffer will switch 
from one logical level to the other. 
It can also be noticed that the resolution depends on the threshold level. 
For the threshold below $200$ mV a~worsening of the quality of time measurement is clearly visible. 
One should add that there is no data delivered from the producer of the FPGA about this type 
of measurements and hence it can be only guessed that it is most likely related to the internal properties 
of LVDS buffers. 
However, the most importantly contribution to the overall measurement precision is very good and should be below $70$ ps RMS when shifting the base line of negative pulses coming from TOF-PET to $2$ V. This is done on the MVT mezzanine board. The $70$ ps RMS time measurement precision is the worst case scenario it deals with a slowest expected J-PET PMT signal. For signals where rise time is above $0.5$ V/ns it reaches $\approx 30$ ps RMS ($\approx 20$ ps RMS per channel). It is close to the intrinsic TRB TDC measurement precision.

It is necessary to emphasize some other consequences of using FPGA LVDS buffers as comparators. When signal is just crossing barely the threshold the precision of measured time is worsen, this applies to all types of comparators but it is more visible in case of an FPGA LVDS buffer. However we are not worried that it will worsen an overall performance of our system. The J-PET system was designed for signals which amplitudes are larger then $200$ mV. Additionally we did observe that if signal is shorter than $\approx 100$ ps there is a possibility that it will not be detected at the LVDS buffer. It is a quite extreme situation and it only applies when signal just barely crosses the threshold. In case of the J-PET detector the signal width amounts at least $2$ ns.

\section{J-PET detector application} 
One of the MVT FPGA basic applications is the measurement of fast signals coming from the plastic scintillators of the TOF-PET detector  
being developed by the J-PET collaboration \cite{JPET_Bio2011,pet_general,TimeRes,JPET_nim2014}. 

\begin{figure}[!t]
\centering
\includegraphics[width=0.4\textwidth]{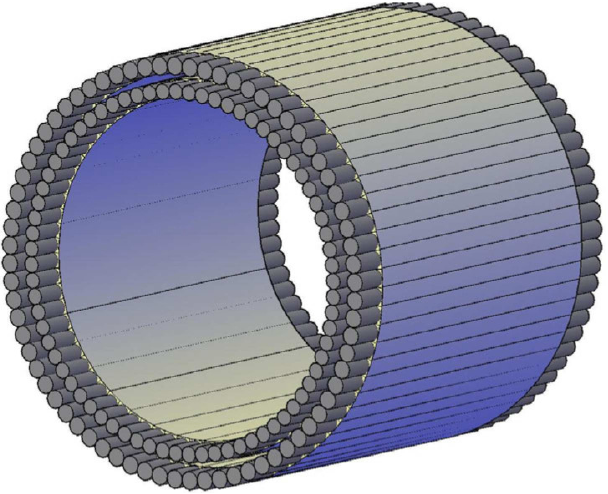}
\caption{The model of two layer version of the J-PET detector.}
\label{jpet_det}
\end{figure}

\begin{figure}[!t]
\centering
\includegraphics[width=0.8\textwidth]{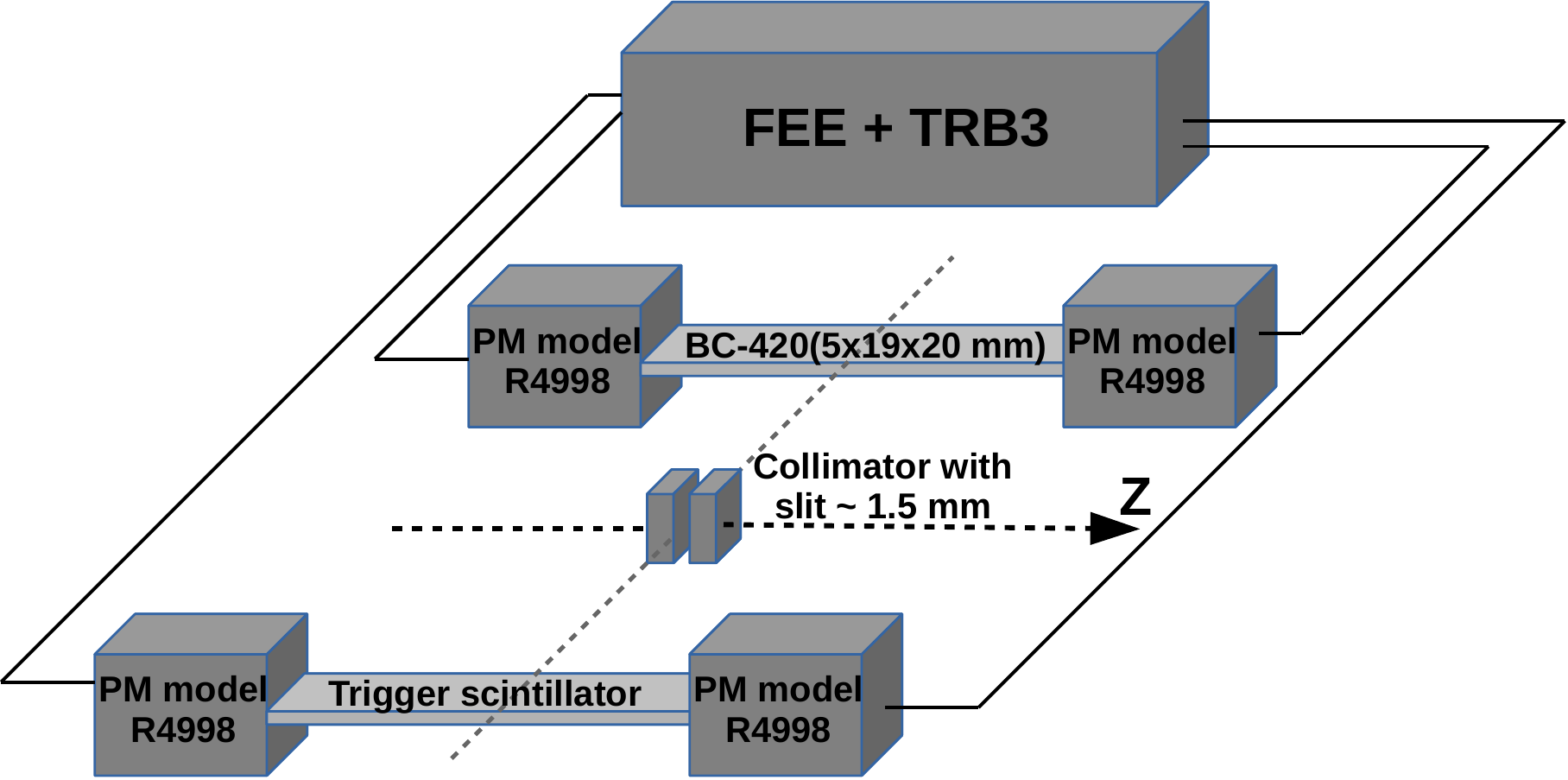}
\caption{Experimental setup used to test the performance of the J-PET detector with the electronic readout described in this article. The detailed description is in the text.}
\label{jpet_schem}
\end{figure}

In Figure \ref{jpet_det} the schematic view of the J-PET detector is shown. 
The J-PET prototype consists of 196 scintillators arranged in the form of two cylindrical layers, with 
a diameter of about $80$ cm \cite{JPET_symulacje, Potential, feasibility} Figure \ref{jpet_det}.
Figure \ref{jpet_schem} shows schematically a two layer version of the possible arrangement of the J-PET detector.
Each scintillator is read-out at both sides by means of photomultipliers (PMT) 
and signals from each PMT are sampled at four different levels (both rising and falling edges). 
Additionally, based on the reconstructed signal shape \cite{lech_nim_2} 
an estimation of deposited energy (proportional to a signal charge) 
in the scintillator will be possible. 
This information will be used to suppress background originating from events where 
the gamma quantum 
are scattered in the patient's body \cite{pet_general}. 

\begin{figure}[!t]
\centering
\includegraphics[width=1\textwidth]{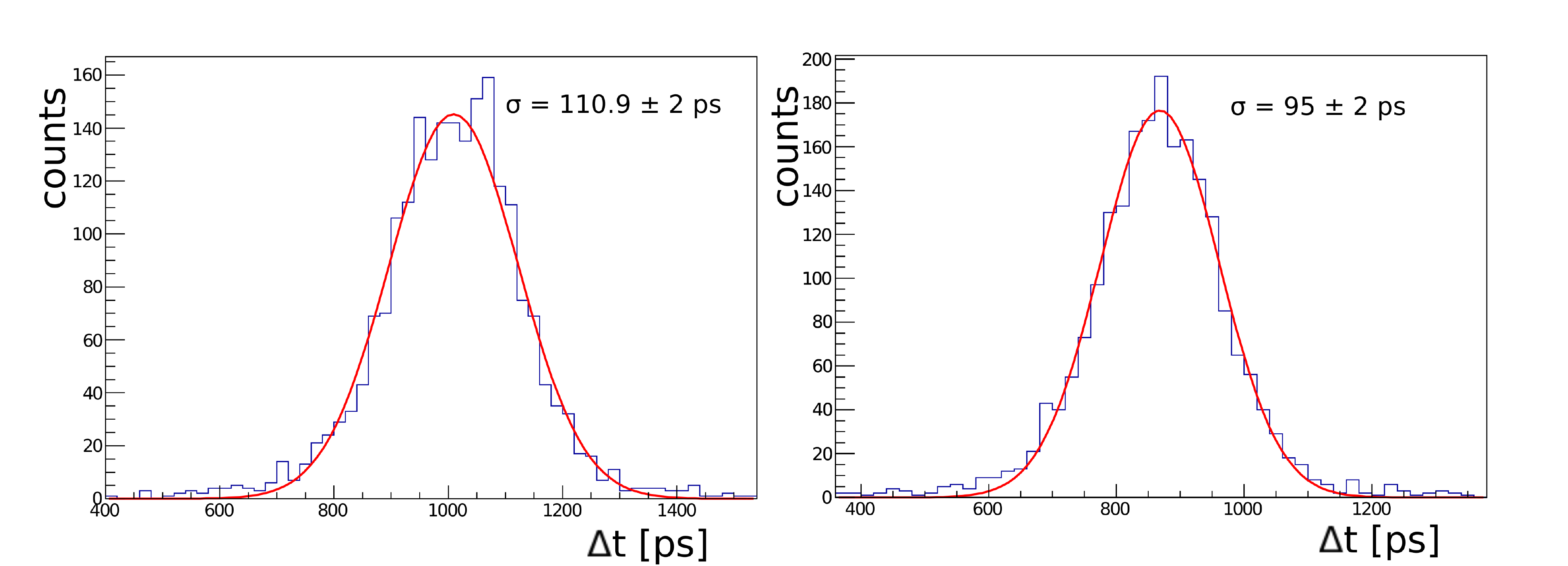}
\caption{Distribution of the difference between times recorded at the lowest threshold for the rising edge of signals from the left and right photomultipliers. Before (left) and after (right) time walk correction.}
\label{pet_resolution_results}
\end{figure}

In order to investigate potential of the presented MVT technique a detector 
set-up shown in Figure \ref{jpet_schem} has been used. 
It consisted of $20$ mm long plastic sintillator bars (BC-420) \cite{SaintGobain}
with cross~section of ($5\times 19$ mm$^2$). 
Both side of the detector have been read-out by PMT tubes from Hamamatsu (R4998) \cite{Hamamatsu}
connected to the electronic readout described in this article. A $^{22}$Na collimated source has been used to illuminate middle part of the detector. Figure \ref{pet_resolution_results} shows distribution of the time difference 
between two signals arriving from both ends of the detector and triggered by the trigger scintillator. 
The triggering and collimation ensured that only $511$ keV gamma quanta, 
relevant for the positron emission tomography, were selected \cite{JPET_nim2014}. 
The times of signals arrival were determined at the rising edge for the lowest applied threshold only.
Thus, it is expected that using time stamps determined at other thresholds should further improve
the time resolution, as discussed above and as described e.g. in reference \cite{lech_nim_2}.
In this work it is demonstrated yet another method to improve a time resolution.
The right figure shows the distribution of the time difference 
%($\Delta t_{corr}$)
after corrections for the walk effect resulting from a time dispersion 
due to the variation of the signal amplitude. Applied corrections were calculated based 
on linear fit to a function of the measured time difference between 
two MVT channels (with the same level of threshold) and the width of the signal determined from time over the threshold. 
The slope of this fit ($a$) was used to recalculate measured 
time: $t_{corrected}=t_{measured}- width \cdot a$.
The achieved resolution of 
$\sigma(\Delta t)=\sigma(t_{left}~-~t_{right})=95$ ps  
implies that the resolution of hit-time determination amounts to 
$\sigma(t_{hit}) = \sigma((t_{left} + t_{right})/2) \approx 48$ ps 
and as a consequence the measurement precision of the time difference between two detectors
is equal to about $\sigma(TOF)=68$ ps. This result proves that a significant improvement of time measurement precision
is possible with respect to the current TOF-PET systems with the best $\sigma(TOF)\approx147$ ps \cite{PETobecnie}.

\section{Conclusion}
In this paper a compact system based on FPGAs 
for the fast signals sampling utilising MVT technique has been presented. 
The measurement sub-system consist of 192 TDC channels which allow to analyse 48 TOF-PET modules signals. 
These by means of measuring a time when PMT signal is crossing four pre-defined thresholds (four for each channel).

Its advantages are: simplified electronic circuit, reduced power consumption, 
low costs, front-end electronics merged with digital electronics 
and more compact final design. 
It has been shown that the combination of FPGA LVDS buffers acting as comparators with the implementation 
of TDC inside FPGA provides a very good performance in terms of the time resolution 
and reconstruction of the signal charge. 
It has been demonstrated that intrinsic MVT TDC channel precision is on the level of $20$ to $70$ ps for a single channel and
depends on a measured signal slope. Additionally it is shown that the charge of the signals can be determined with the fractional precision better than 10\%.
When used as a readout for the developed J-PET detector, it was possible to achieve 
Time-of-Flight resolution of 
$\sigma(TOF)\approx68$ ps which is by about a factor of two better with respect to the
resolution of the current TOF-PET tomography systems \cite{PETobecnie}.

\acknowledgments

The authors acknowledge the technical support 
by A. Heczko, W. Migdal, 
and the financial support from the Polish National Center for Development 
and Research through grant INNOTECH-K1/IN1/64/159174/NCBR/12, 
the EU and MSHE Grant no. POIG.02.03.00-161 00-013/09,
National Science Center Poland based on decision number DEC-2013/09/N/ST2/02180 and UMO-2016/21/B/ST2/01222.

%\end{otherlanguage}
\end{document}